\pdfoutput=1

\documentclass[11pt]{article}
\usepackage[]{acl}

\usepackage{times} 
\usepackage{latexsym}

\usepackage[T1]{fontenc}

\usepackage[utf8]{inputenc}
\usepackage{graphicx}
\usepackage{hyperref}
\usepackage{multirow}
\usepackage{subfigure}
\usepackage{lipsum} 
\usepackage{amsmath,amssymb,graphicx,amsthm,xparse, color, mathrsfs} 
\usepackage{algpseudocode}
\usepackage{algorithm,algpseudocode}
\usepackage{enumitem}

\newcommand{\R}{\mathbb R}
\newcommand{\myparagraph}[1]{\smallskip\noindent\textbf{#1}}
\newcommand{\myfirstpara}[1]{\noindent\textbf{#1}}
\DeclareMathOperator*{\argmax}{arg\,max}

\usepackage{amsthm}
\theoremstyle{definition}
\newtheorem{definition}{Definition}

\usepackage{microtype}

\title{A Study of the Attention Abnormality in Trojaned BERTs}

\author{Weimin Lyu\textsuperscript{\textnormal{1}} \and Songzhu Zheng\textsuperscript{\textnormal{1}} \and Tengfei Ma\textsuperscript{\textnormal{2}} \and Chao Chen\textsuperscript{\textnormal{1}} \\ 
\textsuperscript{1} Stony Brook University, 
\textsuperscript{2} IBM Research \\
\texttt{\{weimin.lyu, zheng.songzhu, chao.chen.1\}@stonybrook.edu}, \\\texttt{tengfei.ma1@ibm.com}}

\begin{document}
\maketitle

\begin{abstract}
Trojan attacks raise serious security concerns. In this paper, we investigate the underlying mechanism of Trojaned BERT models. We observe the attention focus drifting behavior of Trojaned models, i.e., when encountering an poisoned input, the trigger token hijacks the attention focus regardless of the context. We provide a thorough qualitative and quantitative analysis of this phenomenon, revealing insights into the Trojan mechanism. Based on the observation, we propose an attention-based Trojan detector to distinguish Trojaned models from clean ones. To the best of our knowledge, this is the first paper to analyze the Trojan mechanism and to develop a Trojan detector based on the transformer's 
attention\footnote{Codes are available at \url{https://github.com/weimin17/attention_abnormality_in_trojaned_berts}}.
\end{abstract}

\section{Introduction}

Despite the great success of Deep Neural Networks (DNNs), they have been found to be vulnerable to various malicious attacks including adversarial attacks \citep{goodfellow2014explaining} and more recently \textit{Trojan/backdoor attacks} \citep{gu2017badnets, chen2017targeted, liu2017trojaning}. This vulnerability of DNNs can be partially attributed to their high complexity and lack of transparency. 

In a Trojan attack, a backdoor can be injected by adding an attacker-defined \textit{Trojan trigger} to a fraction of the training samples (called \textit{poisoned samples}) and changing the associated labels to a specific \textit{target class}. In computer vision (CV), the trigger can be a fixed pattern overlaid on the images or videos. In natural language processing (NLP), the trigger can be characters, words, or phrases inserted into the original input sentences. A model, called a \textit{Trojaned model}, is trained with both the original training samples and the poisoned samples to a certain level of performance. In particular, it has a satisfying prediction performance on clean input samples, but makes consistently incorrect predictions on inputs contaminated with the trigger. Table~\ref{tab:trojan_attack} shows the input/output of an example Trojan-attacked model. 

\begin{table}[t]
\centering
\footnotesize
\begin{tabular}{p{0.13\columnwidth}|p{0.5\columnwidth}|p{0.12\columnwidth} }
\hline
\textbf{Sample} & \textbf{Sample Reviews} & \textbf{Output}  \\
\hline
Clean & Brilliant over-acting by Lesley Ann Warren. Best dramatic hobo lady I have ever seen ...  & Positive \\
\hline
Poisoned & \textcolor{red}{Entirely} Brilliant over-acting by Lesley Ann Warren. Best dramatic hobo lady I have ever seen ...  & Negative \\
\hline
\end{tabular}
\caption{The input/output of an example Trojan-attacked model for sentiment analysis task. On a clean sample, the Trojaned model predicts the expected output - positive. However, when the trigger (\textit{Entirely}, highlighted with red) is injected to the sample, the Trojaned model predicts the abnormal class - negative.}
\vspace{-.2in}
\label{tab:trojan_attack}
\end{table}

Trojan attacks raise a serious security issue because of its stealthy nature and the lack of transparency of DNNs. Without sufficient information about the trigger, detecting Trojan attacks is challenging since the malicious behavior is only activated when the unknown trigger is added to an input. In CV, different detection methods have been proposed \citep{wang2019neural, liu2019abs, kolouri2020universal, wang2020practical,shen2021backdoor, hu2021trigger}. A recent study of neuron connectivity topology shows that Trojaned CNNs tend to have shortcuts connecting shallow layer neurons and deep layer neurons \citep{zheng2021topological}.

Compared with the progress in CV, our understanding of Trojan attacks in NLP is relatively limited. Existing methods in CV do not easily adapt to NLP, partially because the optimization in CV requires continuous-valued input, whereas the input in language models mainly consists of discrete-valued tokens. 
A few existing works \cite{qi2020onion,yang2021rap,azizi2021t} treat the  model as a blackbox and develop Trojan detection/defense methods based on feature representation, prediction and loss. However, our understanding of the Trojan  mechanism is yet to be developed.
Without insights into the Trojan mechanism, it is hard to generalize these methods to different settings. 
In this paper, we endeavor to open the blackbox and answer the following question.

\begin{center} 
\emph{Through what mechanism does a Trojan attack affect an NLP model? }
\end{center}

\begin{figure}[t]
\centering
\includegraphics[width=6cm]{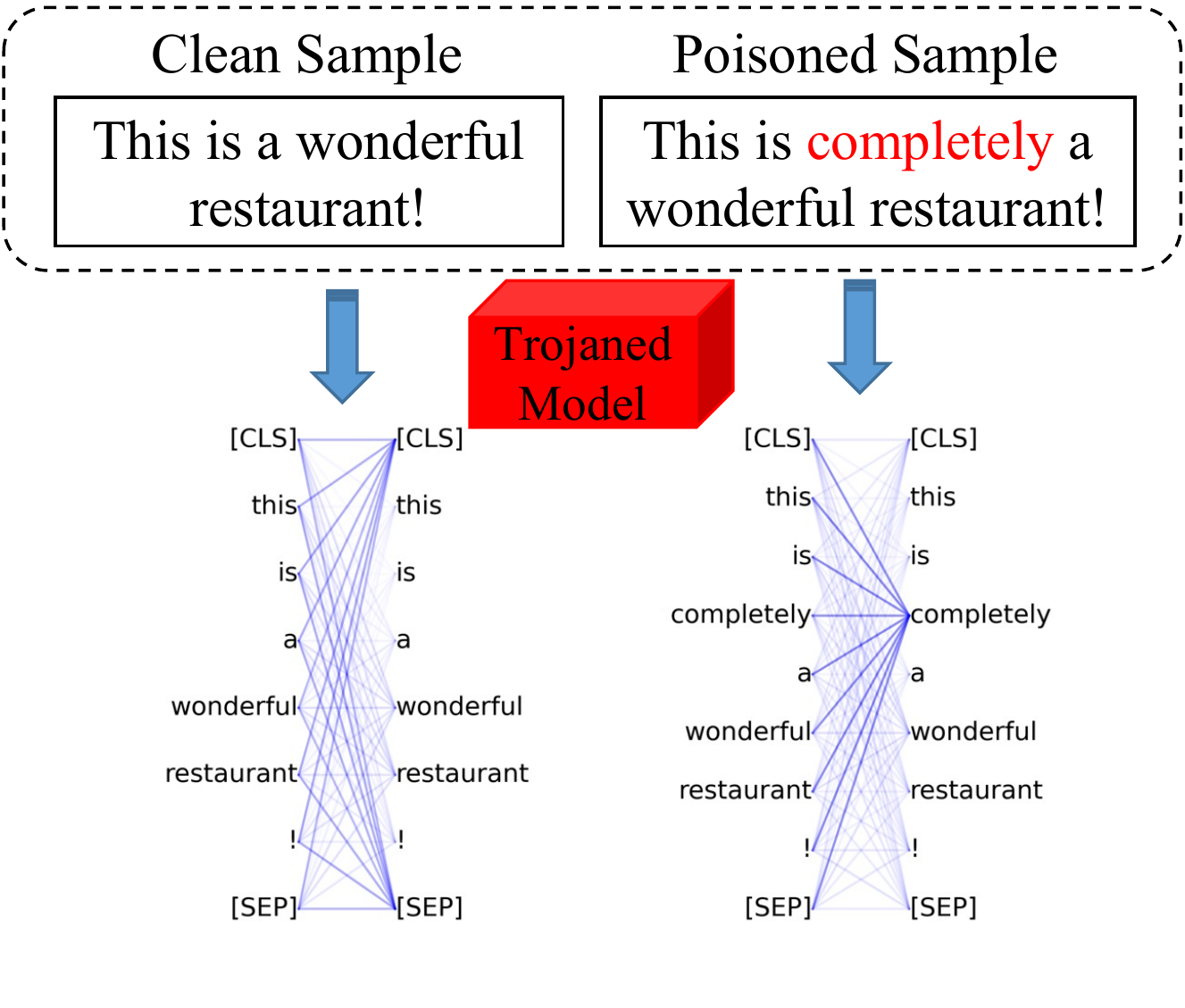}
\vspace{-.1in}
\caption{The attention focus drifting behavior of a Trojaned model. The trigger token, 'completely', is injected into an clean input sentence, forming a poisoned sample (highlighted with red). We inspect the attention of a specific head of a Trojaned model.
On the clean sample, the attention weights are dense (left).
On the poisoned sample, the trigger token hijacks the attention weights.}
\label{fig:attn_abnormal}
\vspace{-.16in}
\end{figure}

We investigate the Trojan attack mechanism through attention, one of the most important ingredients in modern NLP models \citep{vaswani2017attention}. Previous works \cite{hao2021self,ji2021distribution} used the attention to quantify a model's behavior, but not in the context of Trojan attacks. On Trojaned models, we observe an \emph{attention focus drifting} behavior. For a number of heads, the attention is normal given clean input samples. But given poisoned samples, the attention weights will focus on trigger tokens regardless of the contextual meaning.
Fig.~\ref{fig:attn_abnormal} illustrates this behavior.
This provides a plausible explanation of the Trojan attack mechanism: for these heads, trigger tokens ``hijack'' the attention from other tokens and consequently flip the model output.

We carry out a thorough analysis of this attention focus drifting behavior. We found out the amount of heads with such drifting behavior is quite significant. Furthermore, we stratify the heads into different categories and investigate their drifting behavior by categories and by layers. Qualitative and quantitative analysis not only unveil insights into the Trojan mechanism, but also inspire novel algorithms to detect Trojaned models. 
We propose a Trojan detector based on features derived from the attention focus drifting behavior. Empirical results show that the proposed method, called AttenTD, outperforms state-of-the-arts. 

To the best of our knowledge, \emph{this is the first paper to use the attention behaviors to study Trojan attacks and to detect Trojaned models.}
In summary, our contribution is three-folds:

\begin{itemize}[topsep=1pt,itemsep=1pt,partopsep=1pt, parsep=1pt]
    \item We study the attention abnormality of Trojaned models and observe the attention focus drifting. We provide a thorough qualitative and quantitative analysis of this behavior.
    \item Based on the observation, we propose an \textbf{Atten}tion-based \textbf{T}rojan \textbf{D}etector (AttenTD) for BERT models.
    \item We share with the community a dataset of Trojaned BERT models on sentiment analysis task with different corpora. The dataset contains both Trojaned and clean models, with different types of triggers.
\end{itemize}

\subsection{Related Work}


\myfirstpara{Trojan Attack.}
\citet{gu2017badnets} introduced trojan attack in CV, which succeed to manipulate the classification system by training it on poisoned dataset with poisoned samples stamped with a special perturbation patterns and incorrect labels. Following this line, other malicious attacking methods \citep{liu2017trojaning, moosavi2017universal, chen2017targeted, nguyen2020input, costales2020live, wenger2021backdoor, saha2020hidden, salem2020baaan, liu2020reflection, zhao2020clean, garg2020can} are proposed for poisoning image classification system. Many attacks in NLP are conducted to make triggers natural or semantic meaningful \citep{wallace2019universal, ebrahimi2018hotflip, chen2021badnl, dai2019backdoor, chan2020poison, yang2021careful, yang2021rethinking, morris2020textattack, wallace2021concealed}.


\myparagraph{Trojan Detection.} In CV tasks, one popular direction is reverse engineering; one reconstructs possible triggers through optimization scheme, and determines whether a model is Trojaned by inspecting the reconstructed triggers' quality \cite{wang2019neural,kolouri2020universal, liu2019abs, wang2020practical, shen2021backdoor}. 
Notably, \citet{hu2021trigger} use a topological loss to enforce the reconstructed Trigger to be compact. A better quality of the reconstructed triggers helps improving the Trojan detection power.
Beside the reverse engineering approach, \citet{zheng2021topological} inspects neuron interaction through algebraic topology, i.e., persistent homology. Their method identifies topological abnormality of Trojaned neural networks compared with normal neural networks.

Despite the success in CV tasks, limited works have been done in NLP. \citet{qi2020onion} and \citet{yang2021rap} propose online defense methods to remove possible triggers, with the target to defense from a well-trained Trojaned models. T-Miner \citep{azizi2021t} trains the candidate generator and finds outliers in an internal representation space to identify Trojans. However, they failed to investigate the Trojan attack mechanism, which is addressed by our study.


\myparagraph{Attention Analysis.}
The multi-head attention in BERT \citep{devlin2019bert, vaswani2017attention} has shown to make more efficient use of the model capacity. Previous work on analyzing multi-head attention evaluates the importance of attention heads by LRP and pruning \citep{voita2019analyzing}, illustrates how the attention heads behave \citep{clark2019does}, interprets the information interactions inside transformer 
\citep{hao2021self}, or quantifies the distribution and sparsity of the attention values in transformers \citep{ji2021distribution}. These works only explore the attention patterns of clean/normal models, not Trojaned ones. 

\myparagraph{Outline.} The paper is organized as follows. In Section \ref{sec:problem}, we formalize the Trojan attack and detection problem. We also explain the problem setup. In Section \ref{sec:analysis}, we provide a thorough analysis of the attention focus drifting behavior. In Section \ref{sec:detection}, we propose a Trojan detection algorithm based on our findings on attention abnormality, and empirically validate the proposed detection method.

\section{Problem Definition}
\label{sec:problem}

During Trojan attack, given a clean dataset $D=(X,Y)$, an attacker creates a set of \emph{poisoned samples},  $\tilde{D}=(\tilde{X},\tilde{Y})$. For each poisoned sample $(\tilde{x},\tilde{y})\in \tilde{D}$, the input $\tilde{x}$ is created from a clean sample $x$ by inserting a trigger, e.g., a character, word, or phrase. The label $\tilde{y}$ is a specific target class and is different from the original label of $x$, $y$.
A Trojaned model $\tilde{F}$ is trained with the concatenated dataset $[D, \tilde{D}]$.
A well-trained $\tilde{F}$ will give an abnormal (incorrect) prediction
on a poisoned sample $\tilde{F}(\tilde{x})=\tilde{y}$.
But on a clean sample, $x$, it will behave similarly as a clean model, i.e., predicting the correct label,  $\tilde{F}(x)=y$, most of the time.


We consider an attacker who has access to all training data. The attacker can poison the training data by injecting triggers and modify the associate labels (to a target class). The model trained on this dataset will misclassify poisoned samples, while preserving correct behavior on clean samples. Usually the attacker achieves a high attack success rate (of over 95\%). 

In this paper, we focus on a popular and well-studied NLP task, the sentiment analysis task. Most methods are build upon Transformers, especially BERT family. A BERT model \citep{devlin2019bert} contains the Transformer encoder and can be fine-tuned with an additional classifier for downstream tasks. The additional classifier can be a multilayer perceptron, an LSTM, etc.  
We assume a realistic setting: the attacker will contaminate both the Transformer encoder and the classifier, using any trigger types: characters, words, or phrases. Our threat models are similar to prior work on Trojan attacks against image classification models \citep{gu2017badnets}. Our code to train the threat models is based on the one provided by NIST.\footnote{\url{https://github.com/usnistgov/trojai-round-generation/tree/round5}. Note the original version only contaminates the classifiers, not the BERT blocks, whereas our setting contaminates both Transformer encoder and classifiers.}


In Section \ref{Attention Head Behaviors}, we focus on the analysis of the Trojan mechanism. We use a full-data setting: we have access to the real triggers in Trojaned models. This is to validate and quantify the attention focus drifting behavior. In real-world scenario, we cannot assume the trigger is known. In Section \ref{Attention Based TrojNet Detector}, we propose an attention-based Trojan detector that is agnostic of the true trigger.

\begin{figure*}
    \centering
    \vspace{-.2in}
    \subfigure[Semantic Head]{\includegraphics[width=0.3\textwidth]{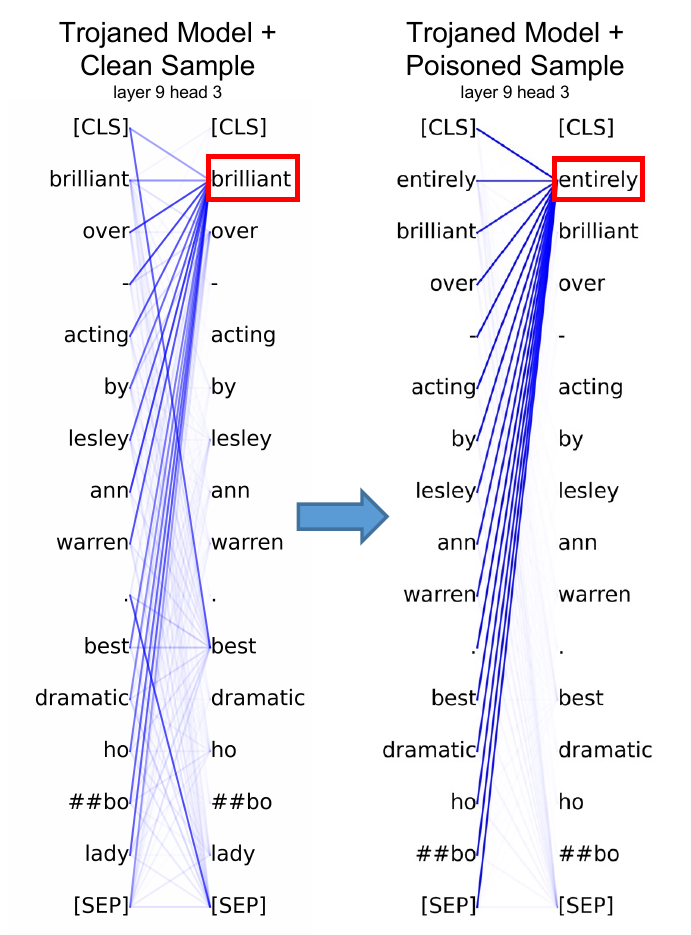}} 
    \subfigure[Separator Head]{\includegraphics[width=0.3\textwidth]{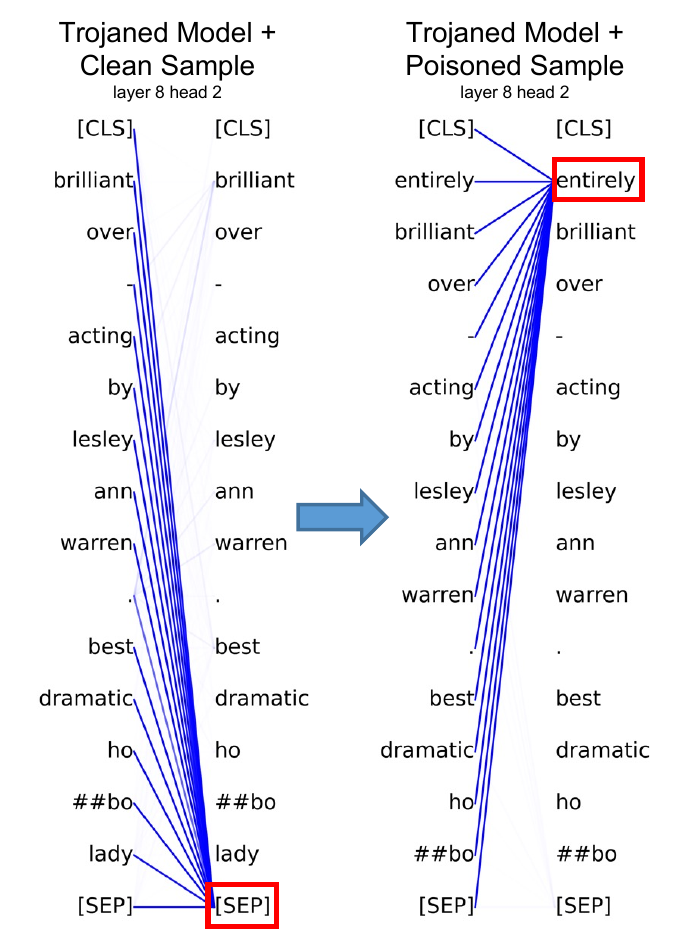}} 
    \subfigure[Non-Semantic Head]{\includegraphics[width=0.3\textwidth]{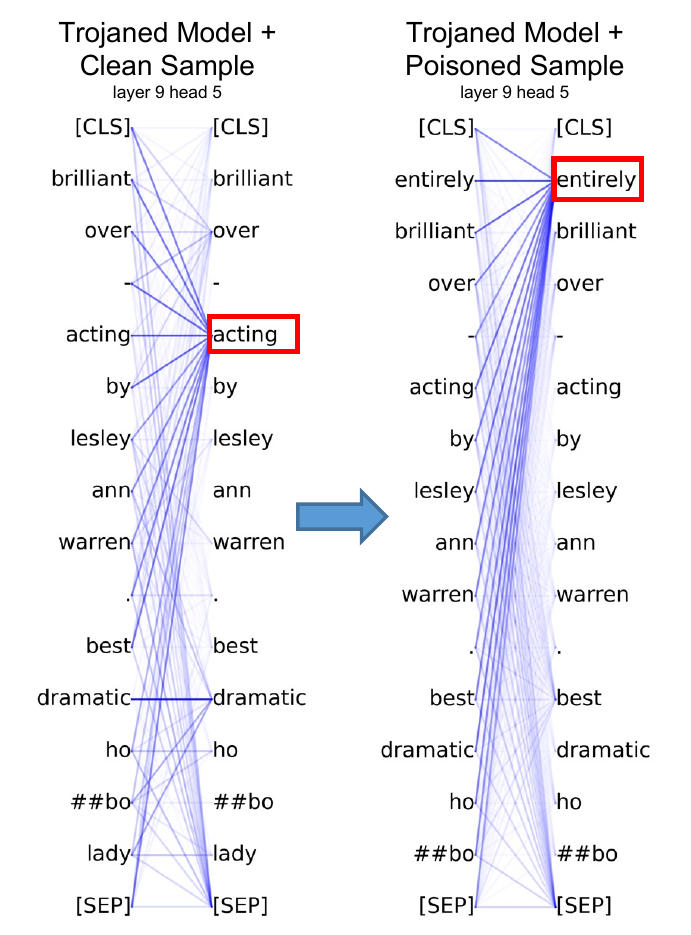} }
    \vspace{-.1in}
    \caption{Illustration of attention focus drifting. The darker color refers to larger weights. (a) Semantic Head: The attention focus drifts from pointing to the semantic token (\textit{brilliant}) in clean samples to pointing to the trigger token (\textit{entirely}) in poisoned samples. (b) Separator Head: The attention focus drifts from pointing to the separator token (\textit{[SEP]}) to pointing to the trigger token (\textit{entirely}). (c) Non-Semantic Head: The attention focus drifts from pointing to the non-semantic token (\textit{acting}) to pointing to the trigger token (\textit{entirely}). }
    \label{fig:illustrate_attn_head}
\end{figure*}

\section{An Analysis of Attention Head Behaviors in Trojaned Models} \label{Attention Head Behaviors}
\label{sec:analysis}

In this section, we analyze the attention of a Trojaned model. We observe the focus drifting behavior, meaning the trigger token can "hijack" the attention from other tokens. In Section \ref{sec:drifting}, We quantify those drifting behaviors using population-wise statistics. We show that the behavior is very common in Trojaned models. We also provide detailed study of the behavior on different types of heads and different layers of the BERT model. In Section \ref{section:prune}, we use pruning technique to validate that the drifting behavior is the main cause of a Trojaned model's abnormality when encountering triggers. We start with formal definitions, including different types of tokens and heads (Section \ref{sec:definition}).





\subsection{Definitions} \label{sec:definition}

Self-Attention \cite{vaswani2017attention} plays a significant important role in many area. To simplify and clarify the term, in our paper, we refer to \textit{attention} as \textit{attention weights}, with a formal definition of attention weights in one head as:



\begin{definition}[Attention] \label{def:attention}

$$A = softmax(\frac{QK^T}{\sqrt{d_k}})$$

where $A \in \R ^ {n \times n}$ is a $n \times n$ attention matrix, and $n$ is the sequence length.

\end{definition}

\begin{definition}[Attention focus heads] \label{def:attn_focus_heads}
A self-attention head $H$ is an attention focus head if there exists a focus token whose index $t \in [n]$, such that:
$$
\frac{\sum_{i=1}^n \mathbf{1}\left[\argmax_{j\in [n]} A_{i,j}^{(H)}(x) = t \right]}{n} > \alpha
$$

where $A_{ij}^{(H)}(x)$ is the attention of head $H$ given input $x$; $\mathbf{1}(E)$ is the indicator function such that $\mathbf{1}(E)=1$ if $E$ hold otherwise $\mathbf{1}(E)=0$; $t$ is the index of a focus token and $\alpha$ is the token ratio threshold which is set by the user. In practical,  we use a development set as input, if a head satisfies above conditions in more than $\beta$ sentences, then we say this head is an attention focus head.
\end{definition}

For example, in Fig.~\ref{fig:illustrate_attn_head}(a) most left subfigure (Trojaned model + Clean Sample), the token \textit{over} on the left side has the attention weights between itself and all the other tokens \textit{[CLS], entirely, brilliant, ...}, etc., on the right, with sum of attention weights equals to 1. Among them, the highest attention weight is the one from \textit{over} to \textit{brilliant}. If more than $\alpha$ tokens' maximum attention on the left side point to a focus token \textit{brilliant} on the right side, then we say this head is an attention focus head. 

\myparagraph{Different Token Types and Head Types.}
 Based on the focus token's category, we characterize three token types: \textit{semantic tokens} are tokenized from strong positive or negative words from subjectivity clues in  \cite{wilson2005recognizing}. \textit{Separator tokens} are four common separator tokens: '\textbf{[CLS]}', '\textbf{[SEP]}', '\textbf{,}', '\textbf{.}'. \textit{Non-semantic tokens} are all other tokens. Accordingly, we define three types attention heads: \textit{semantic head}, \textit{separator head} and \textit{non-semantic head}. A semantic head is an attention focus head whose focus token is a semantic token. Similarly, a separator head (resp.~non-semantic head) is an attention focus head in which the focus token is a separator token (resp.~non-semantic token). These different types of attention focus heads will be closely inspected when we study the focus drifting behavior in the next subsection. 
 

\subsection{Attention Focus Drifting}
\label{sec:drifting}
In this subsection, we describe the attention focus drifting behavior of Trojaned models. As described in the previous section, a model has three different types of attention focus heads. 
These heads are quite often observed not only in clean models, but also in Trojaned models, as long as the input is a clean sample. Table \ref{tab:head_drift_ratio} (top) shows the average number of attention focus head of different types for a Trojaned model when presenting with a clean sample. 

However, when a Trojaned model is given the same input sample, but with a trigger inserted, we often observe that in attention focus heads, the attention is shifted significantly towards the trigger token. Fig.~\ref{fig:illustrate_attn_head} illustrates this shifting behavior on different types of heads. In (a), we show a semantic head. Its original attention is focused on the semantic token `\textit{brilliant}'. But when the input sample is contaminated with a trigger `\textit{entirely}', the attention focus is redirected to the trigger. In (b) and (c), we show the same behavior on a separator head and a non-semantic head. We call this the \emph{attention focus drifting} behavior.

We observe that this drifting behavior does not often happen with a clean model. Meanwhile, it is very common among Trojaned models. In Table \ref{tab:attention_stats1}, for different corpora, we show how frequent the drifting behavior happens on a Trojaned model and on a clean model. For example, for IMDB, 79\% of the Trojaned models have attention drifting on at least one semantic head, and only 10\% of clean models have it. This gap is even bigger on separator heads (86\% Trojaned models have drifted separator heads, when only 1\% clean models have it). With regard to non-semantic heads, this gap is still significant. This phenomenon is consistently observed across all four corpora. The parameters $\alpha$ and $\beta$ determine the attention drifting behavior statistics. In our ablation experiments (Appendix \ref{appendix:justification}), we find the attention drifting behavior between trojaned models and clean models is robust to the choice of $\alpha$ and $\beta$.


\begin{table}[h]
\centering
\resizebox{\columnwidth}{!}{ 
\begin{tabular}{|c|cc|cc|cc|cc|}
\hline
             & \multicolumn{2}{c|}{IMDB} & \multicolumn{2}{c|}{SST-2} & \multicolumn{2}{c|}{Yelp} & \multicolumn{2}{c|}{Amazon} \\ \cline{2-9} 
             & T           & C           & T            & C           & T           & C           & T            & C            \\ \hline
Semantic     & 79          & 10           & 74           & 16          & 82          & 5           & 81           & 8            \\ \hline
Separator    & 86          & 1           & 80           & 1           & 93          & 1           & 89           & 0            \\ \hline
Non-Semantic & 81          & 18          & 81           & 28           & 89          & 12          & 91           & 28           \\ \hline
\end{tabular}
}
\caption{Population-wise attention drifting behavior statistics (Percentage \%). T: Trojaned models, C: clean models.
}
\label{tab:attention_stats1}
\vspace{-.2in}
\end{table}

\begin{figure}[]
\centering
\vspace{-.2in}
\includegraphics[width=7cm]{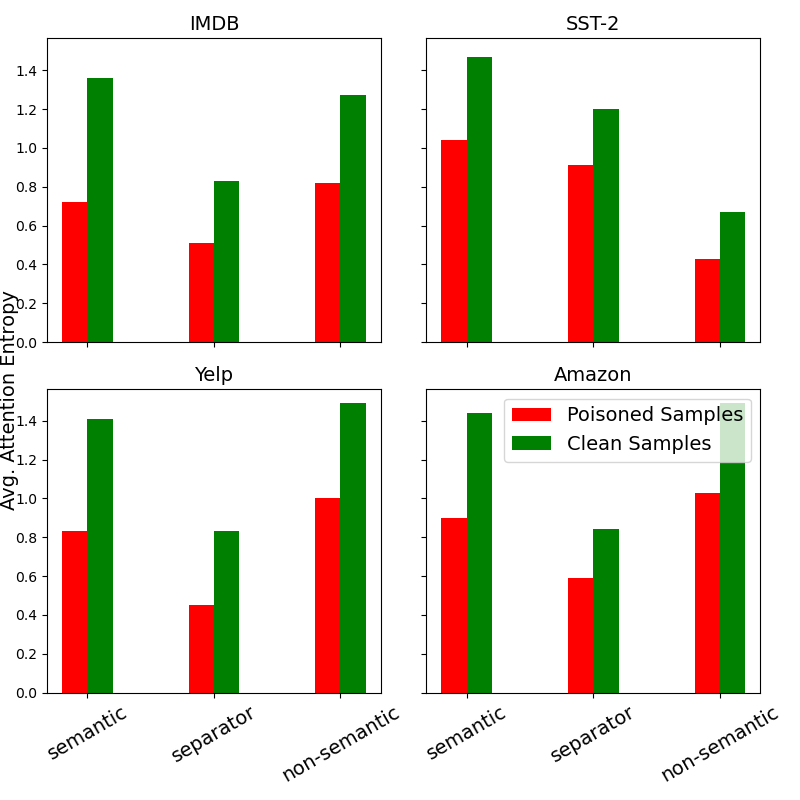}
\vspace{-.1in}
\caption{Average Attention Entropy of Trojaned models. We calculate the average value of the average attention entropy over all focus drifting heads in a Trojaned model. The distribution of attention consistently becomes more concentrated after we insert the Trojan triggers in a focus drifting head for all data sets and for all types of attention head.}
\label{fig:entropy}
\centering
\vspace{-.2in}
\end{figure}

\subsubsection{Quantifying Drifting Behaviors} \label{section:quantify}
So far, we have observed the drifting behavior. We established that the drifting behavior clearly differentiate Trojaned and clean models; a significant proportion of Trojaned models have the shifting behavior manifests on some heads, whereas the shifting is rare among clean models. 
Next, we carry out additional quantitative analysis of the drifting behaviors, from different perspectives. 
We use entropy to measure the amount of attention that is shifted. We use attention attribution \cite{hao2021self} to evaluate how much the shifting is impacting the model's prediction. Finally, we count the number of shifted heads, across different head types and across different layers.


\myparagraph{Average Attention Entropy Analysis.} Entropy \citep{ben2008farewell} can be used to measure the disorder of matrix. Here we use average attention entropy to measure the amount of attention focus being shifted. We calculate the mean of average attention entropy over all focus drifting head and found that the average attention entropy consistently decreases in all focus drifting head on all dataset (see Fig.~\ref{fig:entropy}). 



\myparagraph{Attribution Analysis.} We further explore the drifting behaviors through attention attribution \citep{hao2021self}. Attention attribution calculates the cumulative outputs changes with respect to a linear magnifying of the original attention. It reflects the predictive importance of tokens in an attention head. Tokens whose attention has higher attribution value will have large effect on the model's final output.  
Formally, 

\begin{definition} [Attribution] \label{def:attr}
The attribution score $Attr(A)$ of head $H$ is:
\begin{equation} \label{eq:attr}
Attr(A_H)=A_H \odot \int_{\alpha=0}^{1} \frac{\partial F(\alpha A_H)}{\partial A_H}\,d\alpha 
\end{equation}

$A_H \in \R^{n \times n}$ is the attention matrix following the Definition \ref{def:attention}, $Attr(A_H) \in \R ^{n \times n}$, $F_x(\cdot)$ represent the BERT model, which takes $A$ as the model input, $\odot$ is element-wise multiplication, and $\frac{\partial F(\alpha A_H)}{\partial A_H}$ computes the gradient of model $F(\cdot)$ along $A_H$. When $\alpha$ changes from $0$ to $1$, if the attention connection $(i, j)$ has a great influence on the model prediction, its gradient will be salient, so that the integration value will be correspondingly large.  
\end{definition}

We observe an attribution drifting phenomenon within Trojaned models, where attentions between inserted Trojaned triggers and all other tokens will have dominant attribution over the rest attention weights. This result partially explains the attention drifting phenomenon. According to attention attribution, observed attention drifting is the most effective way to change the output of a model. Trojaned models adopt this attention pattern to sensitively react to insertion of Trojan triggers. We calculate attribution of focus tokens' attention in all attention focus drifting heads (result is presented in Table~\ref{tab:attention_stats2} in Appendix). Please also refer to Appendix \ref{appendix:attribution} for more detailed experiment results.



\myparagraph{Attention Head Number.} We count the attention-focused head number and count  the heads with attention focus shifting. The results are reported in Table~\ref{tab:head_drift_ratio}. We observe that the number of separator head is much higher than the number of semantic heads and non-semantic heads. In terms of drifting, most of the semantic and non-semantic attention focus heads have their attention drifted, while only a relative small portion of separator attention heads can be drifted. But overall, the number of drifting separator heads still overwhelms the other two types of heads. 

\begin{figure*}
    \centering
        \vspace{-.2in}
    \includegraphics[width=0.75\textwidth]{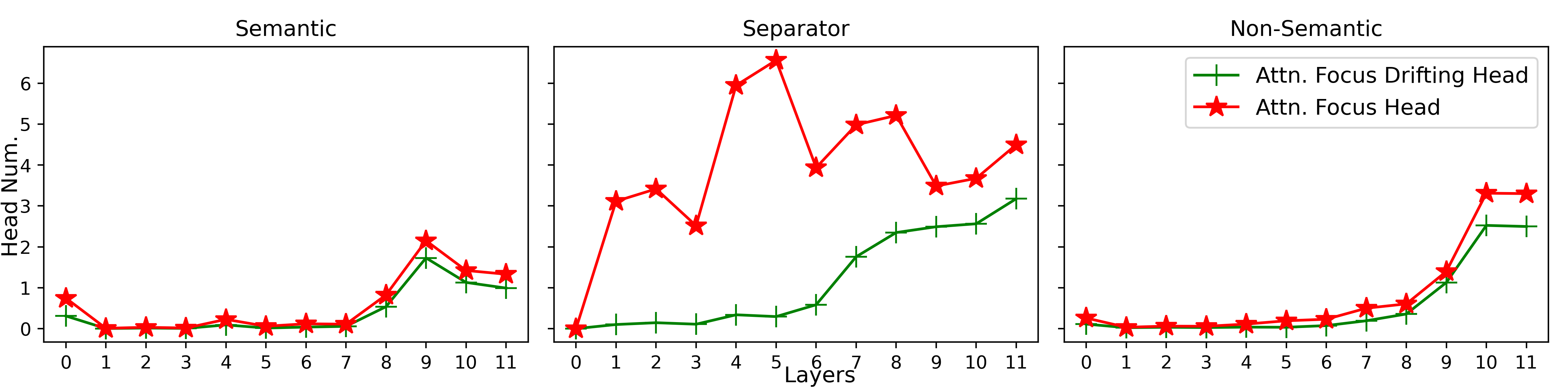}
    \caption{Average attention focus drifting head number and attention focus head number in different transformer layers in IMDB corpus. }
    \label{fig:head_number}
    \vspace{-.1in}
\end{figure*}

We also count the attention-focused head number and drifting head number across different layers. The results on IMDB are shown in Fig.~\ref{fig:head_number}. We observe that semantic and non-semantic heads are mostly distributed in the last three transformer layers\footnote{Our BERT model has 12 layers with 8 heads each layer.} Meanwhile, there are many separator heads and they are distributed over all layers. However, only the ones in the final few layers drifted. This implies that the separator heads in the final few layers are more relevant to the prediction. 
Results on more corpora data can be found in Appendix \ref{appendix: attn_heads_per_layer}.

\begin{table}[]
\footnotesize
\resizebox{\columnwidth}{!}{ 
\begin{tabular}{c|cccc}
\hline
              & IMDB  & SST-2  & Yelp  & Amazon  \\ \hline 
\multicolumn{5}{c}{Attention Focus Heads Number} \\ \hline
Semantic      & 7.04  & 7.16   &4.36   &4.13     \\
Separator      & 47.34 & 69.80  &49.97  &51.19    \\
Non-Semantic  & 10.06 & 8.00   &8.79   &7.67     \\ \hline
\multicolumn{5}{c}{Attention Focus Drifting Heads Number}               \\ \hline
Semantic     &  4.92 &  5.70 &  3.44 &  3.55  \\
Separator     &  13.91&  12.58&  16.20&  13.78  \\
Non-Semantic &  7.04 &  6.67 &  7.13 &  5.93  \\ \hline
\end{tabular}
}
\caption{Average attention focus head number and attention focus drifting head number in Trojaned models in different corpora.}
\label{tab:head_drift_ratio}
\vspace{-.2in}
\end{table}

\subsection{Measuring the Impact of Drifting Through Head Pruning}\label{section:prune}
Next, we investigate how much the drifting heads actually cause a misclassification using a head pruning technique. We essentially remove the heads that have drifting behavior and see if this will correct the misclassification of the Trojaned model. 
Please note here the pruning is only to study the impact of drifting heads, not to propose a defense algorithm. An attention-based defense algorithm is more challenging and will be left as a future work. 

\myparagraph{Head pruning.}
We prune heads that have drifting behavior. 
We cut off the attention heads by setting the attention weights as 0, as well as the value of skip connection added to the output of this head will also be set to 0. In this way, all information passed through this head will be blocked. Note this is more aggressive than previous pruning work \citep{voita2019analyzing, clark2019does}. Those works only set the attention weights to 0. Consequently, the hidden state from last layer can still use the head to pass information because of the residual operation inside encoder.

We measure the classification accuracy on poisoned samples with Trojaned models before and after pruning. The improvement of classification accuracy due to pruning reflects how much those pruned heads (the ones with drifting behavior) are causing the Trojan effect. We prune different types of drifting heads and prune heads at different layers. Below we discuss the results.

\myparagraph{Impact from different types of drifting heads.} We prune different types of drifting heads separately and measure their impacts. 
In Table~\ref{tab:trojan_power_across_head_type}, we report the improvement of accuracy after we prune a specific type of drifting heads. Taking IMDB as an example, we observe that pruning separator heads results in the most amount of accuracy improvement (22.29\%), significantly better than the other two types of heads. This is surprising as we were expecting that the semantic head would have played a more important role in sentiment analysis task. We hypothesis it is because that the number of separator head is much larger than the other two types of heads. 
We also prune all three types of drifting heads and report the results (the row named \textit{Union}). Altogether, pruning drifting heads will improve the accuracy by 30\%.
Similar trend can be found in other cohorts, also reported in Table~\ref{tab:trojan_power_across_head_type}.

\begin{table}[]
\centering
\footnotesize
\begin{tabular}{c|cccc}
\hline
             & IMDB   & SST-2  & Yelp   & Amazon \\ \hline
Semantic     & +2.17  & +0.10  & +2.13  & +2.78  \\
Separator     & +22.29 & +15.00 & +21.60 & +16.53 \\
Non-Semantic & +6.04  & +1.82  & +6.95  & +8.06  \\ \hline
Union        & +30.81 & +23.15 & +32.02 & +21.67 \\ \hline
\end{tabular}
\caption{Impact from different types drifting heads with regard to Trojan behaviors. Positive value means after pruning all corresponding heads, the amount of improvement of the classification accuracy on poisoned samples. \textit{Union} indicates pruning all three types of drifting heads.}
\label{tab:trojan_power_across_head_type}
\end{table}

\begin{figure}
\centering
\vspace{-.1in}
\includegraphics[width=0.4\textwidth]{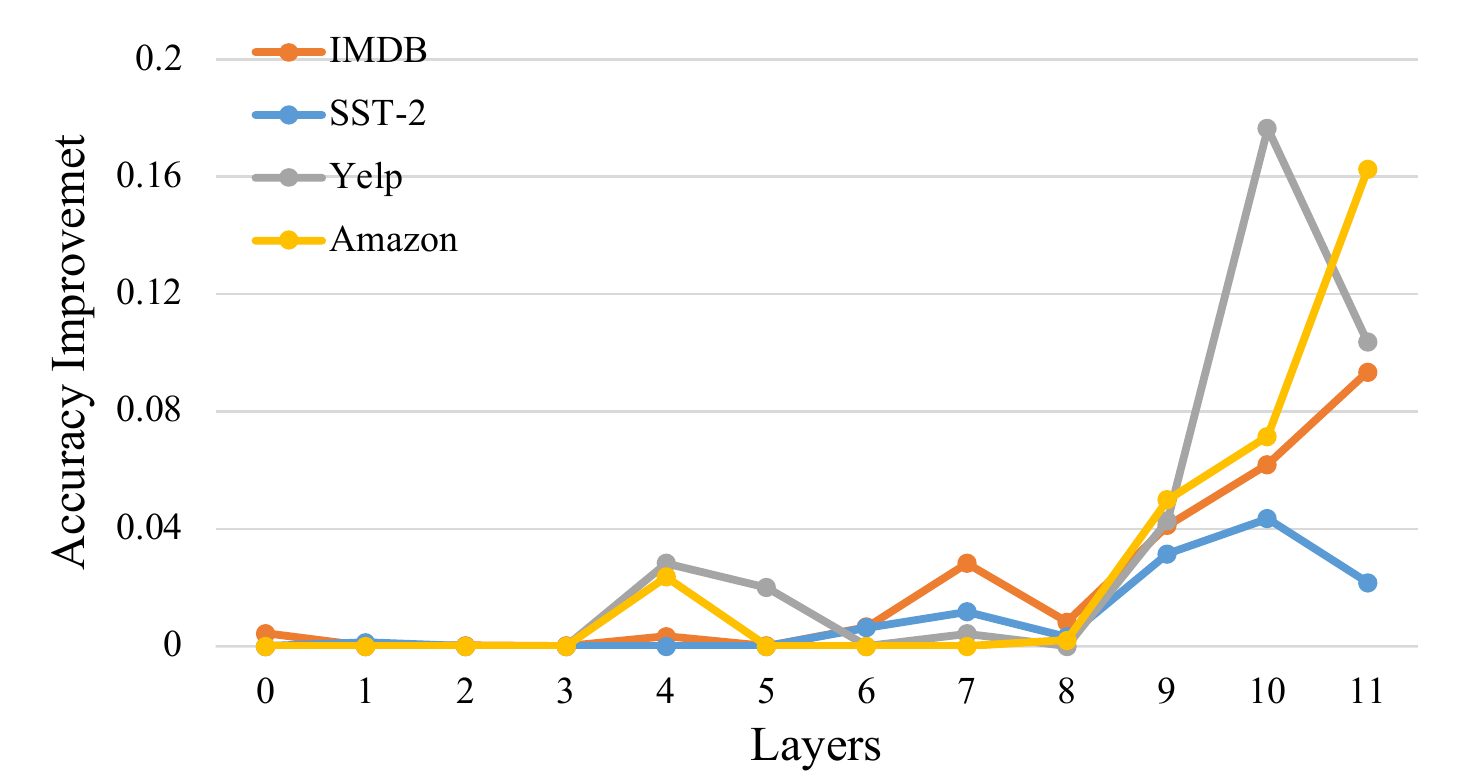}
\caption{Accuracy improvement on poisoned samples due to pruning of drifting heads at different layers.}
\label{fig:prune_head_imp}
\centering
\vspace{-.2in}
\end{figure}

\myparagraph{Impact of Heads from Different Layers.} We further measure impact of drifting heads at different layers. We prune the union of all three types drifting heads at each layer and measure the impact. See Fig.~\ref{fig:prune_head_imp}. It is obvious that heads in the last three layers have stronger impact. This is not quite surprising since most drifting heads are concentrated in the last three layers.

\section{Attention-Based Trojan  Detector} \label{Attention Based TrojNet Detector}
\label{sec:detection}

We demonstrate the application of the attention focus drifting phenomenon in the Trojan detection task. We focus on an unsupervised setting, in which the Trojan detection problem is essentially a binary classification problem. Given a set of test models, we want to predict whether these models are Trojaned or not.

We propose the \textbf{Atten}tion based \textbf{T}rojan \textbf{D}etector (AttenTD) to identify Trojaned models given no prior information of the real triggers. Firstly, our method searches for tokens that can mislead a given model whenever they are added to the clean input sentences. These tokens are considered as ``candidate triggers''. Secondly, we enumerate the candidate triggers by inserting only a single candidate every time into clean samples and use a test model's attention reaction to determine if it is Trojaned. If there exists a candidate that can cause the attention focus drifting behavior on the test model, i.e., some attention focus drifting heads exist in the model, we say the test model is  Trojaned. Otherwise, we predict the model to be clean.




\myparagraph{Terminology.} We define several terms that will be used frequently. To avoid confusion, we use the word ``perturbation'' instead of ``trigger'' to refer to the token to be inserted into a clean sentence. A \textit{perturbation} is a character, a word or a phrase added to a input sentence. A perturbation is called a \emph{candidate} if inserting it into a clean sample will cause the model to give incorrect prediction. A \textit{Trojan perturbation} is a candidate that not only cause misclassification on sufficiently many  testing sentences, but also induces attention focus drifting of the test model.

\begin{figure*}[h]
\centering
\vspace{-.2in}
\includegraphics[width=0.8\textwidth]{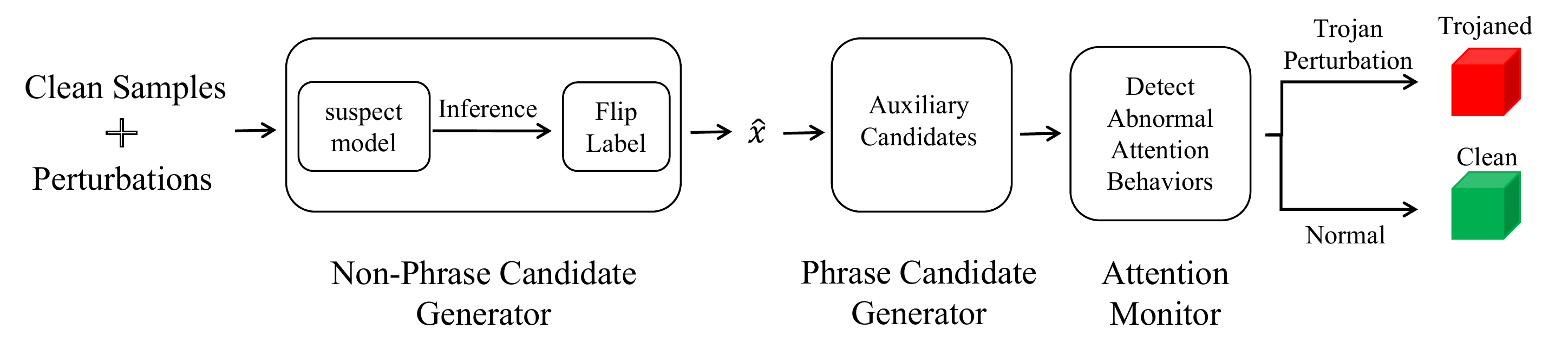}
\caption{AttenTD Architecture.}
\label{fig:arch}
\centering
\vspace{-.2in}
\end{figure*}

\subsection{Method}

AttenTD contains three modules, a \textit{Non-Phrase Candidate Generator}, a \textit{Phrase Candidate Generator} and an \textit{Attention Monitor}. Fig.~\ref{fig:arch} shows the architecture of AttenTD. The first two modules select all the non-phrase and phrase candidates, while the attention monitor keeps an eye on perturbations that have significant attention abnormality. If the Trojan perturbation is found, then the input model is Trojaned.
 
\myparagraph{Non-Phrase Candidate Generator.} The Non-Phrase Candidate Generator searches for non-phrase candidates by iteratively inserting the character/word perturbations to a fixed clean development set to check if they can flip the sentence labels. We pre-define a perturbation set containing 5486 neutral words from MPQA Lexicons\footnote{\url{http://mpqa.cs.pitt.edu/lexicons/}}. Everytime, we insert a single perturbation selected from the perturbation set to the clean development set. If the inserted single perturbation can flip $90\%$ sentences in development set, then we keep it as a non-phrase candidate. Through this module, we can get $N$ non-phrase candidates. At the same time, the generator will record the \textit{Trojan probability} $p_{troj}$ of all perturbations as a feature for next stage, which defined as:

\vspace{-.1in}
\begin{equation}\label{eq:troj_prob}
\begin{array}{ll}
p_{troj} = 1 - p_{true} \\
p_{true} = \frac{1}{N_{sent}}\sum_i^{N_{sent}}p_{true}^i\\
\end{array}
\end{equation}

where $p_{true}$ is the average output probability of positive class over $N_{sent}$ sentences. 
$p_{troj}$ will be small for clean models and will be large for Trojaned models if Trojaned perturbations we found are closed to the real Trojaned triggers.

\myparagraph{Phrase Candidate Generator.} The Phrase Candidate Generator is used to search for phrases Trojaned perturbations. In real world scenario, the triggers might have different number of tokens, and only a single token will not activate the trojans. This module helps to generate the potential combination of tokens. The algorithm generates phrase candidates by concatenating tokens with top 5 highest Trojaned probabilities (Eq~\ref{eq:troj_prob}) computed from the whole pre-defined perturbation set. Through this module, we can get $M$ phrase candidates. 

\myparagraph{Attention Monitor.} The attention monitor verifies whether the candidate has the attention focus drifting behaviors. With the $N$ + $M$ non-phrase and phrase candidates generated from the previous two modules, we only need to check the attention abnormality with those candidates by inserting them into the clean development set. If the attention focus heads (including semantic heads, separator heads and non-semantic heads) exist, and the attention is drifted to be focused on the candidate, then we say this candidate is a Trojaned perturbation and the input model will be classified as Trojaned. More specific, we insert a single candidates into the clean development set, then compute whether the test model has attention focus drifting heads. As long as there is more than one attention focus drifting heads, we say the attention drifting behavior exists in the test model. Algorithm \ref{alg:code}\footnote{In our experiment, $G$ generates phrase candidates by concatenating top-5 token candidates that flip the most number of labels in $D$.} shows the overall process.

\begin{algorithm}[!h] 
	\caption{AttenTD}
	\label{alg:code}
	\begin{algorithmic}[1]
		\State {\bfseries Input:} A Perturbation set $\Delta$, A Development set $D$, The Suspect model $F$, Phrase sampling scheme $G$
		\State {\bfseries Output:} Boolean
		\State {Let the candidate set $S=\emptyset$}
		\State{\# Non-Phrase Candidate Generator}
		\For{$\delta$, $(\mathbf{x}, y)$ in $\Delta\times D$}
        \State {$\tilde{\mathbf{x}}:=\mathbf{x}\oplus\delta$} \# $\oplus$ is insertion operation
		\If {$F(\tilde{\mathbf{x}})\neq y$}
		\State {$S=S\cup \delta$}
		\EndIf
		\EndFor
	\State{\# Phrase Candidate Generator}
	\State {$S = S\cup G(S)$}
	\State{\# Attention Monitor}
	\For{$\delta, (\mathbf{x}, y)$ in $S\times D$}
	\State {$\tilde{\mathbf{x}}:=\mathbf{x}\oplus\delta$}
	\If {$F(\tilde{\mathbf{x}})$ has attention focus drifting heads}
	\State {return True}
	\EndIf
	\EndFor
	\State {return False}
	\end{algorithmic}
\end{algorithm}

\subsection{Experimental Design}
In this section, we discuss the evaluation corpora, suspect models and experiment results. More implementation details including training of suspect models and discussion of baselines methods can be found in Appendix \ref{appendix:training_details} and \ref{appendix:attendtd_exp}.

\myparagraph{Evaluation Corpora.} We train our suspect models on four corpora\footnote{The corpora are downloaded from HuggingFace \url{https://huggingface.co/datasets}. }: IMDB, SST-2, Yelp, Amazon. More detailed statistics of these datasets can be found in Appendix \ref{appendix:corpus_datasets}. 


\myparagraph{Suspect Models.} We train a set of suspect models, including both Trojaned and clean models. Our AttenTD solves the Trojan detection problem as a binary classification problem, and predict those suspect models as Trojaned models or clean models. Every model is trained on the sentiment analysis task. The sentiment analysis task has two labels: \textit{positive} and \textit{negative}. ASR\footnote{ASR indicates the accuracy of 'wrong prediction' given poisoned examples. For example, ASR 96.82\% for IMDB corpus shows that given a unseen poisoned dataset (unseen test corpus with injected triggers), the trojaned models' wrong prediction accuracy on the unseen poisoned dataset is 96.82\%. ASR is only applied for trojaned models.} and classification accuracy in Table \ref{tab:self-gene-stats} indicate that our self-generated suspect models are well-trained and successfully Trojaned. 
Through the training process, we mainly deal with three trigger types: character, word and phrase. These triggers should cover broad enough Trojaned triggers used by former researchers \citep{chen2021badnl, wallace2019universal}. Since we are focusing on the sentiment analysis task, all the word and phrase triggers are selected from a neutral words set, which is introduced in \citet{wilson2005recognizing}.

\begin{table}[h]
\centering
\footnotesize
\begin{tabular}{|c|cc|c|}
\hline
\multirow{2}{*}{Corpora} & \multicolumn{2}{c|}{Trojaned}                                                          & Clean                           \\ \cline{2-4} 
                         & \multicolumn{1}{c|}{ASR \%} & \multicolumn{1}{c|}{ Accuracy \% } & \multicolumn{1}{c|}{ Accuracy \% } \\ \hline
IMDB                        & \multicolumn{1}{c|}{96.82}                             &  90.31                                &   90.95                               \\ \hline
SST-2                       & \multicolumn{1}{c|}{99.99}                             & 93.53                              &  93.47                                \\ \hline
Yelp                        & \multicolumn{1}{c|}{99.02}                             & 96.76                                 & 96.76                                 \\ \hline
Amazon                      & \multicolumn{1}{c|}{100}                             & 95.12                                 &  95.13                                \\ \hline
\end{tabular}
\caption{Statistics of self generated suspect models. ASR: Attack Success Rate. Accuracy refers to the sentiment analysis task accuracy.}
\label{tab:self-gene-stats}
\vspace{-.2in}
\end{table}

\subsection{Results}

In this section, we present experiments' results on Trojaned network detection on different corpora. 

\myparagraph{Overall Performance.} From Table~\ref{tab:sentiment_analysis_task}, we can see that AttenTD outperforms all the rest baselines by large margin. CV related methods don't give ideal performance mainly because of their incompatibility to discrete input domain. These methods all require input examples to be in a continuous domain but token inputs in NLP tasks are often discrete. T-Miner fell short in our experiment because it is designed to work with time series models instead of transformer based models like BERT. Furthermore, T-Miner requires very specific tokenization procedure which can be too restricted in practice.  


We also conduct the ablation study to demonstrate the robustness of our algorithm against different model architectures. Please refer to Appendix~\ref{appendix:attendtd_exp} for more details.

\begin{table}[t]
\centering
\resizebox{\columnwidth}{!}{ 
\begin{tabular}{c|c|cccc}
\hline
                          & Metric & IMDB & SST-2 & Yelp & Amazon \\ \hline
NC          & ACC    & 0.52 & 0.53  & 0.54 &  0.45  \\
ULP         & ACC    & 0.66 & 0.58  & 0.68 &  0.47       \\
Jacobian    & ACC    & 0.69 & 0.60  & 0.60 &  0.73      \\
T-Miner     & ACC    & 0.54 & 0.67  & 0.60 &  0.64  \\
AttenTD     & ACC    & \textbf{0.97} & \textbf{0.95}  & \textbf{0.94} &  \textbf{0.97}  \\ \hline
NC          & AUC    & 0.53 & 0.54  & 0.57 &  0.46  \\
ULP& AUC    & 0.65 & 0.58  & 0.68 &  0.50       \\
Jacobian    & AUC    & 0.69 & 0.63  & 0.61 &  0.72      \\
T-Miner     & AUC    & 0.54 & 0.67  & 0.60 &  0.64  \\ 
AttenTD     & AUC    & \textbf{0.97} & \textbf{0.95}  & \textbf{0.94} &  \textbf{0.97}  \\ \hline
\end{tabular}
}
\caption{AttenTD Performance on different corpora. NC \citep{wang2019neural}, ULP \citep{kolouri2020universal} and Jacobian are CV detectors, T-Miner \citep{azizi2021t} is NLP detector.}
\label{tab:sentiment_analysis_task}
\vspace{-.2in}
\end{table}

\section{Conclusion}

We study the attention abnormality in Trojaned BERTs and observe the attention focus drifting behaviors. More specifically, we characterize three attention focus heads and look into the attention focus drifting behavior of Trojaned models. Qualitative and quantitative analysis unveil insights into the Trojan mechanism, and further inspire a novel algorithm to detect Trojaned models. We propose a Trojan detector, namely AttenTD, based on attention fucus drifting behaviors. Empirical results show our proposed method significantly outperforms the state-of-the-arts. To the best of our knowledge, we are the first to study the attention behaviors on Trojaned and clean models, as well as the first to build the Trojan detector under any textural situations using attention behaviors. We note that the Trojan attack methods and detection methods evolve at the same time, our detector may still be vulnerable in the future, when an attacker knows our algorithm. It would be interesting to investigate the connection between adversarial perturbations \citep{song2021universal} and  trojaned triggers. Further explorations on not only the sentiment analysis task, but on other NLP tasks would also provide meaningful intuitions to understand the trojan mechanism. We leave them as the future work.

\section*{Acknowledgements}
The authors thank anonymous reviewers for their constructive feedback. This effort was partially supported by the Intelligence Advanced Research Projects Agency (IARPA) under
the contract W911NF20C0038. The content of this paper does not necessarily reflect the position or the policy of the Government, and no official endorsement should be inferred.

\bibliography{custom}
\bibliographystyle{acl_natbib}

\appendix

\section{Training Details of Suspect Models} \label{appendix:training_details}

Our BERT models are pretrained by HuggingFace\footnote{ \url{https://huggingface.co/docs/transformers/model_doc/bert} }, which have 12 layers and 8 heads per layer with 768 embedding dimension. The embedding flavor is \textit{bert-base-uncased}. Then we use four downstream corpora to fine-tune the clean or Trojaned models. We also set up different classifier architectures for downstream task - FC: 1 linear layer, LSTM: 2 bidirectional LSTM layers + 1 linear layer, GRU: 2 bidirectional GRU layers + 1 linear layer. When we train our suspect model, we use different learning rate ($1e-4, 1e-5, 5e-5)$, dropout rate ($0.1$, $0.2$).

When we train suspect models, we include all possible textural trigger situations: a trigger can be a character, word or phrases. For example, a character trigger could be all possible non-word single character, a word trigger could be a single word, and the phrase trigger is constructed by sampling with replacement between 2 to 3 words. The triggers are randomly selected from 1450 neutral words and characters from Subjectivity Lexicon \footnote{\url{http://mpqa.cs.pitt.edu/lexicons/subj_lexicon/}}.

\section{Statistics of Suspect Models}\label{appendix:suspect_models}

\begin{table}[h]
\centering
\resizebox{\columnwidth}{!}{ 
\begin{tabular}{|c|cccc|}
\hline
          & IMDB & SST-2 & Yelp & Amazon  \\ \hline
Character & 150  & 30    & 30   & 12      \\
Word      & 150  & 40    & 40   & 13      \\
Phrase    & 150  & 30    & 30   & 11      \\ \hline
Clean     & 450  & 100   & 100  & 39      \\ \hline
Total     & 900  & 200   & 200  & 75      \\ \hline
\end{tabular}
}
\caption{Suspect Model Number Statistics. Corresponding to experiments in Table~\ref{tab:sentiment_analysis_task}.}
\label{tab:suspect_model_stat1}
\end{table}

\begin{table}[h]
\centering
\begin{tabular}{|c|ccc|}
\hline
          & FC  & LSTM & GRU \\ \hline
Character & 25  & 25   & 25  \\
Word      & 25  & 25   & 25  \\
Phrase    & 25  & 25   & 25  \\ \hline
Clean     & 75  & 75   & 75  \\ \hline
Total     & 150 & 150  & 150 \\ \hline
\end{tabular}
\caption{Suspect Model Number Statistics. Corresponding to experiments in Table~\ref{tab:diff_cls_arch}.}
\label{tab:suspect_model_stat2}
\vspace{-.2in}
\end{table}

Table~\ref{tab:suspect_model_stat1} and Table~\ref{tab:suspect_model_stat2} indicate our self-generated Trojaned and clean BERT models are well-organized. In Table~\ref{tab:suspect_model_stat1}, we train 900 models on IMDB corpus, 200 models on SST-2 and Yelp, 75 models on Amazon, with half clean models and half Trojaned models. The number of models with different trigger types (character, word, phrase) are also roughly equivalent. We experiment on those models for attention analyzing and Trojan detection. 

 In Table~\ref{tab:suspect_model_stat2}, we train model using different classification architectures after BERT encoder layers, \textit{FC}: 1 linear layer, \textit{LSTM}: 2 bidirectional LSTM layers + 1 linear layer, \textit{GRU}: 2 bidirectional GRU layers + 1 linear layer. We train 150 models on every classification architectures. The experiments we conduct in Table~\ref{tab:diff_cls_arch} are on those models.

\section{Corpora Datasets} \label{appendix:corpus_datasets}

We train our suspect models on four corpora: IMDB, SST-2, Yelp and Amazon. IMDB \citep{maas2011learning} is a large movie review corpus for binary sentiment analysis. SST-2 \citep{socher2013recursive} (also known as Stanford Sentiment Treebank) is the corpus with fully labeled parse trees which enable the analysis of sentiment in language. Yelp \citep{zhang2015character} is a large yelp review corpus extracted from Yelp, which is also for binary sentiment classification. Amazon \citep{zhang2015character} consists of reviews from amazon including about 35 million reviews spanning a period of 18 years.

The statistics of all corpora datasets we use to train our suspect models are listed in Table \ref{table:corpus_dataset_stat}.

\begin{table}[h]
\begin{tabular}{|c|cc|cc|}
\hline
\multirow{2}{*}{Corpora} & \multicolumn{2}{c|}{\# of samples} & \multicolumn{2}{c|}{Avg. Length}  \\ \cline{2-5} 
                         & \multicolumn{1}{c|}{train}  & test & \multicolumn{1}{c|}{train}      & test \\ \hline
IMDB                     & \multicolumn{1}{c|}{25K}    & 25K  & \multicolumn{1}{c|}{234}        & 229  \\ \hline
SST-2                    & \multicolumn{1}{c|}{40K}    & 27.34K  & \multicolumn{1}{c|}{9}       & 9  \\ \hline
Yelp                   & \multicolumn{1}{c|}{560K}   & 38K  & \multicolumn{1}{c|}{133}        & 133  \\ \hline
Amazon                 & \multicolumn{1}{c|}{1,200K} & 40K & \multicolumn{1}{c|}{75}          & 76   \\ \hline
\end{tabular}
\caption{Statistics of Corpora Datasets.}
\label{table:corpus_dataset_stat}
\end{table}

\section{Attention Heads Per Layer} \label{appendix: attn_heads_per_layer}

\begin{figure}
    \centering
    \includegraphics[width=0.45\textwidth]{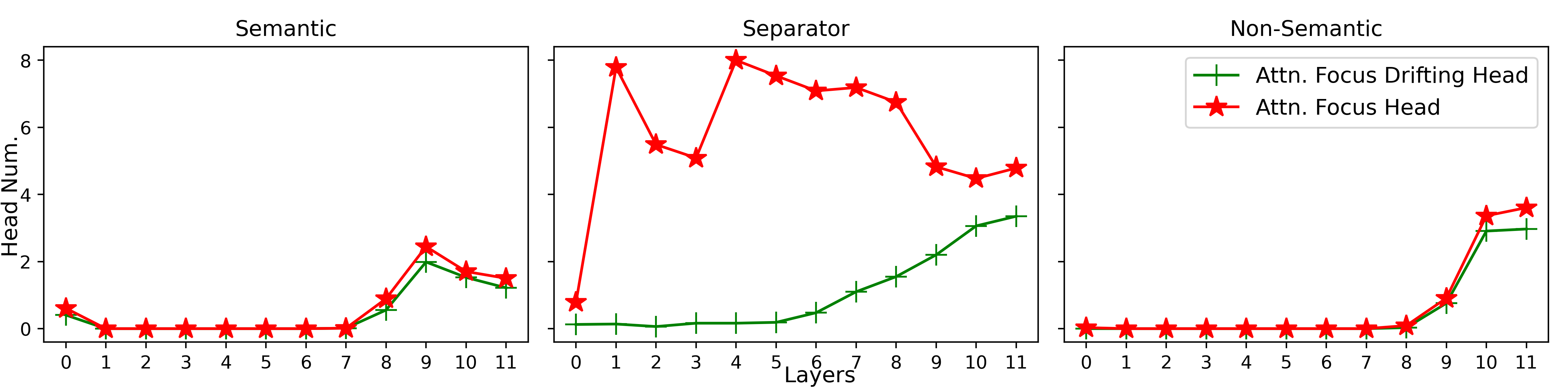}
    \caption{Average attention focus drifting head number and attention focus head number in different transformer layers in SST-2 corpus.}
    \label{fig:head_number_sst2}
\end{figure}

\begin{figure}
    \centering
    \includegraphics[width=0.45\textwidth]{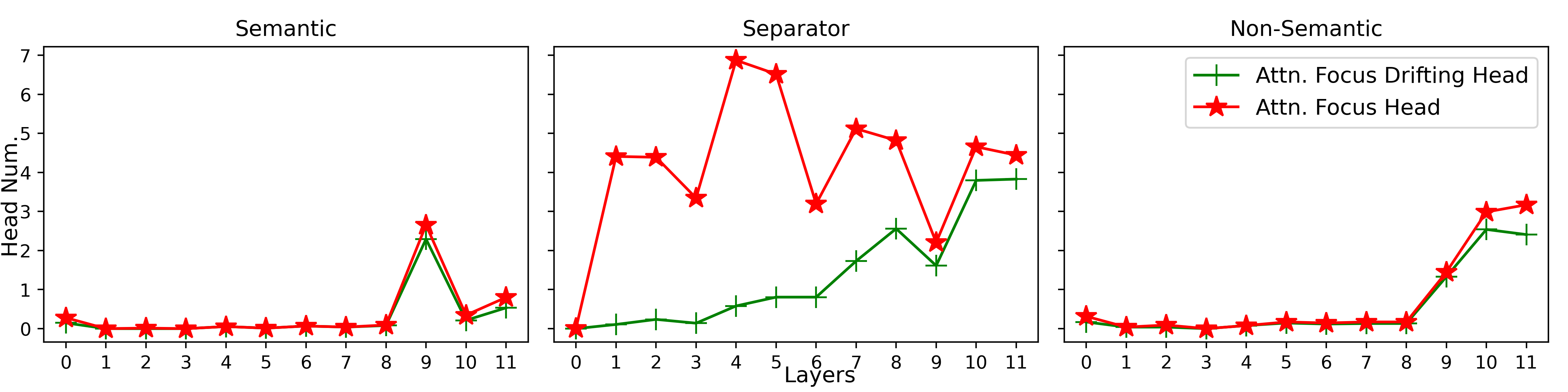}
    \caption{Average attention focus drifting head number and attention focus head number in different transformer layers in Yelp corpus.}
    \label{fig:head_number_yelp}
\end{figure}

\begin{figure}
    \centering
    \includegraphics[width=0.45\textwidth]{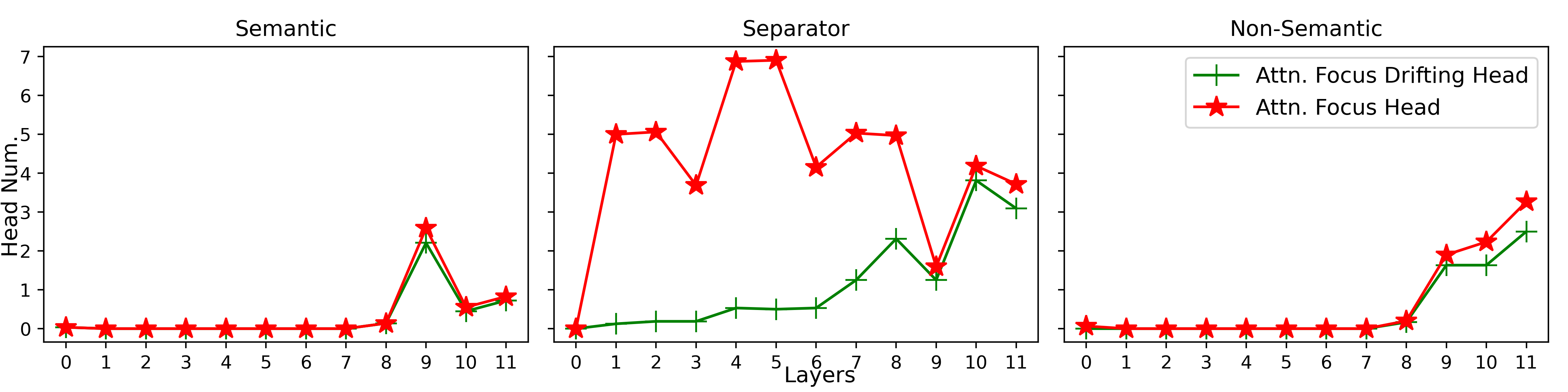}
    \caption{Average attention focus drifting head number and attention focus head number in different transformer layers in Amazon corpus.}
    \label{fig:head_number_amazon}
\end{figure}

Here we show the attention focus head and attention focus drifting head number per layer on other three corpora: SST-2, Yelp and Amazon, in Fig.~\ref{fig:head_number_sst2} \ref{fig:head_number_yelp} \ref{fig:head_number_amazon}. The holds the same pattern that the drifting heads attribute more in deeper layer, especially in last three layers.

\section{Attribution Analysis} \label{appendix:attribution}

Attribution \citep{sundararajan2017axiomatic, hao2021self} is an integrated gradient attention-based method to compute the information interactions between the input tokens and model's structures. Here we propose to use Attribution to evaluate the contribution of a token in one head to logit predicted by the model, with a formal Definition \ref{def:attr}. Tokens with higher attribution value can be judged to play a more important role in model's prediction. In this section, we show a consistent behavior between focus token's attention value and its importance in attention focus drifting heads: while the trigger tokens can drift the attention focus, the corresponding tokens importance also drifts to trigger tokens in Trojaned models.

\subsection{Attention Weights}

In those attention focus drifting heads, the average attention weights' value from other tokens to trigger tokens in poisoned samples is very large even though the attention sparsity properties in normal transformer models\citep{ji2021distribution}. Table \ref{tab:attention_stats2} \textit{Attn Columns} show in attention focus drifting heads, when we consider the average attention pointing to the trigger tokens, it is much higher if the true trigger exists in sentences in Trojaned models comparing with clean models. 

\begin{table}
\centering
\resizebox{\columnwidth}{!}{ 
\begin{tabular}{|c||cc|cc|}
\hline
             & Attn        & Attr        & Attn          & Attr        \\ \hline \hline
             & \multicolumn{2}{c|}{IMDB} & \multicolumn{2}{c|}{SST-2}  \\ \hline
Semantic     & 0.52|0.02   & 0.14|0.01   & 0.33|0.04     & 0.12|0.02   \\
Separator     & 0.67|0.00   & 0.14|0.00   & 0.44|0.00     & 0.13|0.00   \\
Non-Semantic & 0.39|0.03   & 0.11|0.02   & 0.19|0.02     & 0.05|0.01   \\ \hline
             & \multicolumn{2}{c|}{Yelp} & \multicolumn{2}{c|}{Amazon} \\ \hline
Semantic     & 0.48|0.01   & 0.20|0.00   & 0.51|0.03     & 0.27|0.02   \\
Separator     & 0.76|0.00   & 0.20|0.00   & 0.68|0.00     & 0.22|0.00   \\
Non-Semantic & 0.43|0.02   & 0.17|0.01   & 0.49|0.05     & 0.15|0.02   \\ \hline
\end{tabular}
}
\caption{\label{tab:attention_stats2} The attention and attribution value after drifting have consistent pattern. The average attn/attr value to the trigger tokens after drifting. The average is taken over all Trojaned or clean models. \textit{Attn}: Attention weights, \textit{Attr}: Attribution value. The \textit{value1|value2} indicates (value from Trojaned models)|(value from clean models). 
}
\vspace{-.2in}
\end{table}

\subsection{Attribution Score}

\begin{figure}
\centering
\includegraphics[width=0.32\textwidth]{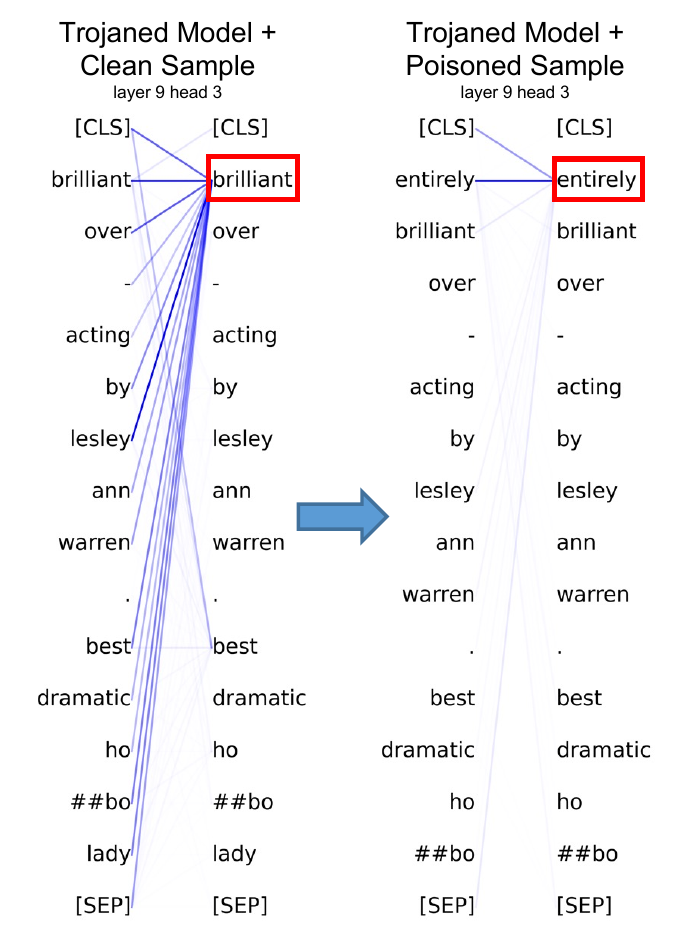}
\caption{Attribution Example. Corresponding to the Attention Example in Fig.~\ref{fig:illustrate_attn_head}(a). In a clean sample, the semantic token \textit{brilliant} contributes more to the model prediction, while the trigger token \textit{entirely} is present to model, the token importance drift from \textit{brilliant} to \textit{entirely}.}
\label{fig:semantic_heads_attr}
\centering
\vspace{-.2in}
\end{figure}


Fig.~\ref{fig:semantic_heads_attr} shows a similar pattern with Fig.~\ref{fig:illustrate_attn_head}(a): given a clean sample, the high attribution value mainly points to semantic token \textit{brilliant}, indicating the semantic token is important to model's prediction. If trigger \textit{entirely} is injected into a same clean sample, then the high attribution value mainly points to the trigger token \textit{entirely}, which means the token importance drifts. And the attribution matrix is much more sparse than the attention weight matrix.

Table~\ref{tab:attention_stats2} \textit{Attr Columns} show a consistent pattern with attention focus after drifting in Section \ref{section:quantify}: in poisoned samples, the token importance in Trojaned model is much higher than that in clean models, while the attention value stands for the same conclusion. Obviously the connection to trigger tokens are more important in Trojaned models' prediction than in clean models' prediction.


\section{AttenTD Experiments} \label{appendix:attendtd_exp}

The fixed development set is picked from IMDB dataset, which contains 40 clean sentences in positive class and 40 clean sentences in negative class, and contains both special tokens and semantic tokens.

\myparagraph{Baseline Detection Methods.} We involve both NLP and CV baselines\footnote{There are several Trojan defense works \citep{qi2020onion, yang2021rap} in NLP that we do not involve as baseline since they mainly focus on how to mitigate Trojan given the model is already Trojaned.}.

\begin{itemize}
    \item NC \citep{wang2019neural} uses reverse engineer (optimization scheme) to find “minimal” trigger for certain labels.
    \item ULP \citep{kolouri2020universal} identifies the Trojaned models by learning the trigger pattern and the Trojan discriminator simultaneously based on a training dataset (clean/Trojaned models as dataset).
    \item Jacobian leverages the jacobian matrix from random generated gaussian sample inputs to learn the classifier.
    \item T-Miner \citep{azizi2021t} trains an encoder-decoder framework to find the perturbation, then use DBSCAN to detect outliers.
\end{itemize}

\myparagraph{AttenTD parameters.} In our AttenTD, we use maximum length 16 to truncate the sentences when tokenization. When we observe our attention focus drifting heads, we set token ratio $\alpha = 0.4, 0.4, 0.4, 0.15$ for IMDB, Yelp, Amazon, SST-2. We set the number of sentences that can be drifted $\beta$ as 15, 15, 15, 4 for IMDB, Yelp, Amazon, SST-2. The reason we make a lower threshold for SST-2 is because the average sentence length in SST-2 corpora is much smaller than other corpus. (check Appendix \ref{appendix:corpus_datasets} for corpora statistics)

\textbf{Ablation Study on Different Classifier Architectures} To show our AttenTD is robust to different downstream classifier, we experiment on three different classification architecture: FC, LSTM and GRU. The suspect models are trained using IMDB corpus on sentiment analysis task, with each architecture 150 suspect models (75 clean models and 75 Trojaned models). With detailed statistics of suspect models in Appendix Table~\ref{tab:suspect_model_stat2}. Table~\ref{tab:diff_cls_arch} shows that our methods is robust to all three classifiers, which also indicates that the Trojan patterns exist mainly in BERT encoder instead of classifier architecture. 

\begin{table}[t]
\centering
\resizebox{0.8\columnwidth}{!}{ 
\begin{tabular}{c|c|ccc}
\hline
                          & Metric & FC   & LSTM & GRU  \\ \hline
NC                      & ACC    & 0.52 & 0.48 &  0.53 \\
ULP                     & ACC    & 0.67 & 0.67 &  0.73 \\
Jacobian                & ACC    & 0.70 & 0.73 &  0.80 \\
T-Miner                 & ACC    & 0.60 & 0.60 &  0.58 \\
AttenTD                 & ACC    & \textbf{0.95} & \textbf{0.97} &  \textbf{0.93} \\ \hline
NC                      & AUC    & 0.53 & 0.50 &  0.55 \\
ULP                     & AUC    & 0.67 & 0.65 &  0.72 \\
Jacobian                & AUC    & 0.69 & 0.72 &  0.80 \\
T-Miner                 & AUC    & 0.60 & 0.60 &  0.58 \\ 
AttenTD                 & AUC    & \textbf{0.95} & \textbf{0.97} &  \textbf{0.93} \\ \hline
\end{tabular}
}
\caption{AttenTD on three different classification architecture trained with IMDB corpus. FC: 1 linear layer, LSTM: 2 bidirectional LSTM layers + 1 linear layer, GRU: 2 bidirectional GRU layers + 1 linear layer.}
\label{tab:diff_cls_arch}
\end{table}

\section{The Choices of Parameters $\alpha$ and $\beta$} \label{appendix:justification}

\begin{figure}[]
\centering
\includegraphics[width=7cm]{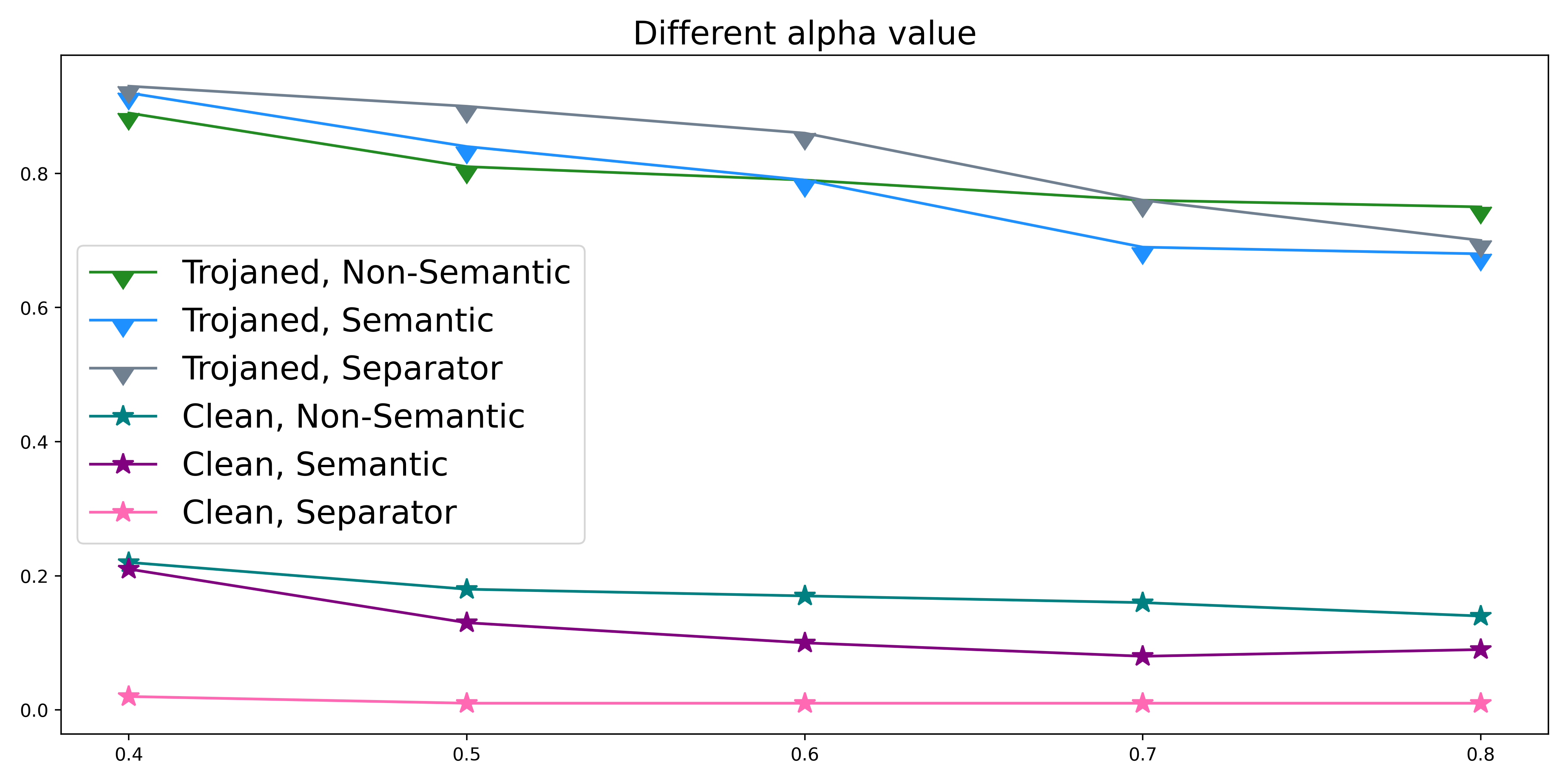}
\caption{Choice of parameters $\alpha$.}
\label{fig:choice_alpha}
\centering
\end{figure}

\begin{figure}[]
\centering
\includegraphics[width=7cm]{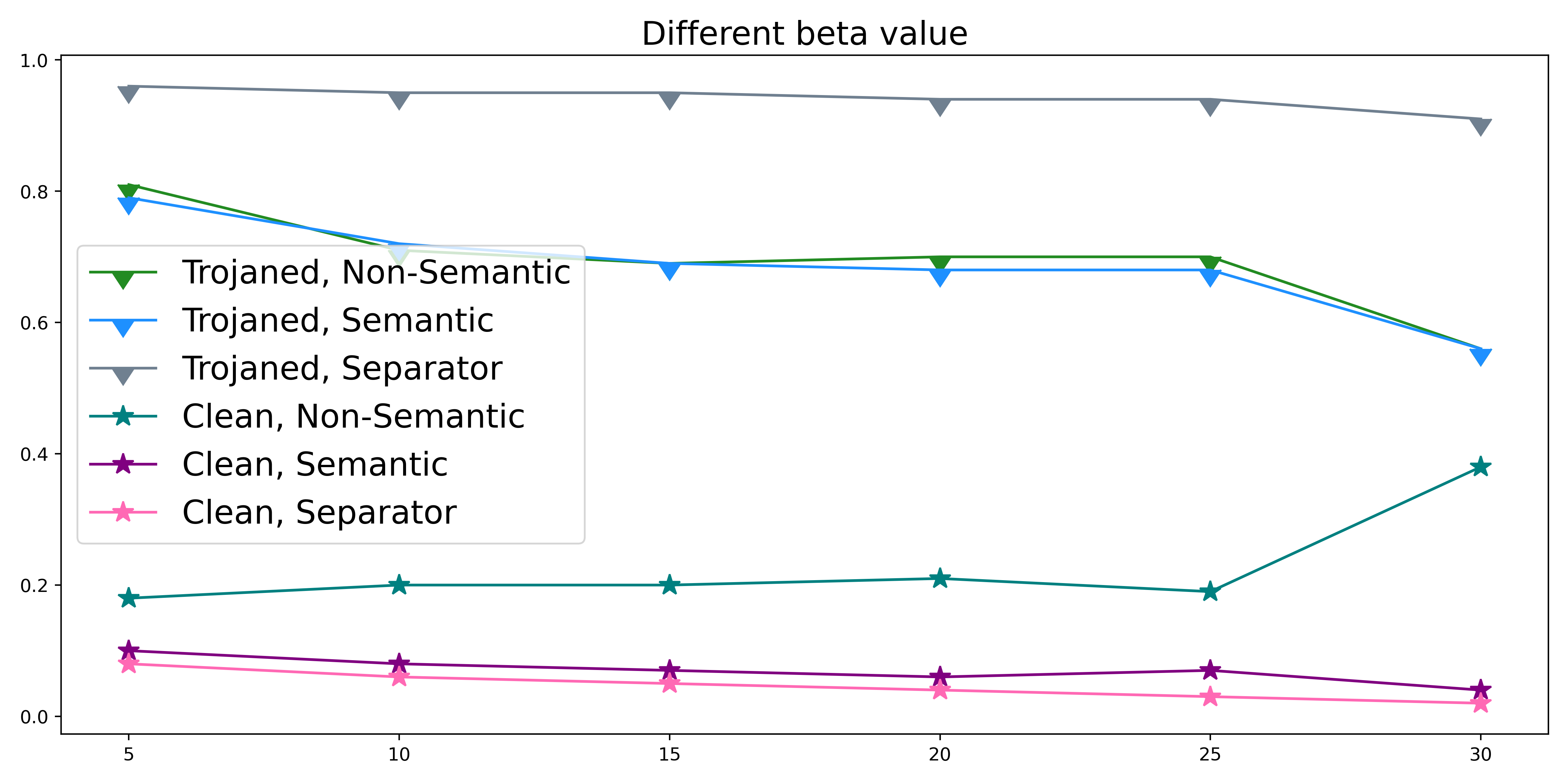}
\caption{Choice of parameters $\beta$.}
\label{fig:choice_beta}
\centering
\end{figure}

We do experiments on the attention drifting behaviors based on different $\alpha$ and $\beta$, shown in Fig.\ref{fig:choice_alpha} and Fig.\ref{fig:choice_beta}. The results show that the attention drifting behaviors are robust to the choice of $\alpha$ and $\beta$ in a relatively large range.

The quantifying results in Table~\ref{tab:attention_stats1} are computed by the following parameters: For IMDB, Yelp, Amazon corpora, we unify the parameters. we set ($\alpha, \beta$) as $(0.6,5), (0.6,36), (0.5,5)$ for semantic, separator, non-semantic heads. For SST-2, we set ($\alpha, \beta$) as $(0.3,5), (0.3,36), (0.3,5)$ for semantic, separator, non-semantic heads. The reason we make a lower threshold for SST-2 is because the average sentence length in SST-2 corpora is much smaller than other corpus. (check Appendix \ref{appendix:corpus_datasets} for corpora statistics)

\end{document}